\documentclass[11pt]{article}
\usepackage{blois,epsfig}

\def\pt{$p_T$}
\def\met{$E_T^{miss}$}

\def\be{\begin{equation}}
\def\ee{\end{equation}}
\def\bea{\begin{eqnarray}}
\def\eea{\end{eqnarray}}

\begin{document}
\vspace*{4cm}
\title{\boldmath{$W/Z$} + jet Production at the LHC}

\author{Gavin Hesketh, on behalf of the ATLAS and CMS collaborations}

\address{Department of Physics and Astronomy, University College London,\\ Gower Street, London, WC1E 6BT, UK}

\maketitle\abstracts{
This paper summarises results on $W$ and $Z$ plus jet production in $pp$ collisions at $\sqrt{s} = 7$~TeV at the CERN Large Hadron Collider, from both the ATLAS and CMS experiments. Based on the 2010 and 2011 datasets, measurements have been made of numerous cross sections provide excellent tests of the latest predictions from QCD calculations and event generators.
}

\section{Introduction}
The leptonic (specifically electron and muon) decay modes of the $W$ and $Z$ (V) boson provide 
very clean experimental signatures
in the complex environment of a hadron collider. At the CERN LHC, the collision energy allows
 enough phase space to produce several high energy QCD jets in association with the $W$ or $Z$ boson,
 and studies of these complex final states provide an excellent testing ground for theoretical
predictions. The variety of possible scale choices (including the boson mass and jet momenta) 
necessitate the use of explicit matrix elements for each additional jet, rather than producing 
jets simply through a parton shower model. While such multi-leg matrix elements have
been calculated at leading order (LO) in perturbation theory for some time, recent theoretical 
developments have increased the reach of next-to-leading order (NLO) calculations out to V 
+ 5 jets, as available at parton level in BLACKHAT~\cite{blackhat}. In terms of full event 
generators, ALPGEN~\cite{alpgen}, MADGRAPH~\cite{madgraph} and SHERPA~\cite{sherpa} match multi-leg LO matrix elements to
parton showers. Complementary efforts to interface NLO matrix elements to parton shower models, such as 
MC@NLO~\cite{mcatnlo} and POWHEG~\cite{powheg} which utilise PYTHIA~\cite{pythia} 
and HERWIG~\cite{herwig} make this a very active area of theoretical development, with several
possible tools to test against the data. As well as being clean calibration sources, V + jets 
also constitute the main source of background in many searches for new physics. A detailed
understanding of these final states is therefore essential.
The ATLAS~\cite{ATLAS} and CMS~\cite{CMS} experiments have performed a variety of cross section 
measurements at a centre of mass energy of 7 TeV, from inclusive quantities such as the rate of
 V+ $\ge1$ jet, through to more 
exclusive quantities like  differential distributions for jets of specific flavours. This paper 
summarises the results obtained to date with the 2010 (approximately 36~pb$^{-1}$) and 2011 
(approximately 5~fb$^{-1}$) datasets, as presented at the Rencontres du Blois, May 2012.

\section{V + inclusive jet production}
At both ATLAS and CMS, the basic $W$ or $Z$ event selection is based on triggering on and 
reconstructing a high transverse momentum 
(\pt) lepton (typically \pt$>20$~GeV), and in the case of the $W$, missing transverse energy (\met)
corresponding to the undetected neutrino (typically \met$>25$~GeV).
Both experiments use the anti-k$_{t}$ algorithm to reconstruct jets, albeit with different radius 
parameter settings ($R=0.4$ at ATLAS, 0.5 at CMS).
Cross sections are generally presented within a fiducial volume, and corrected to the level 
of particles entering the detector, to minimise dependence on theoretical corrections.

\begin{figure}
\psfig{figure=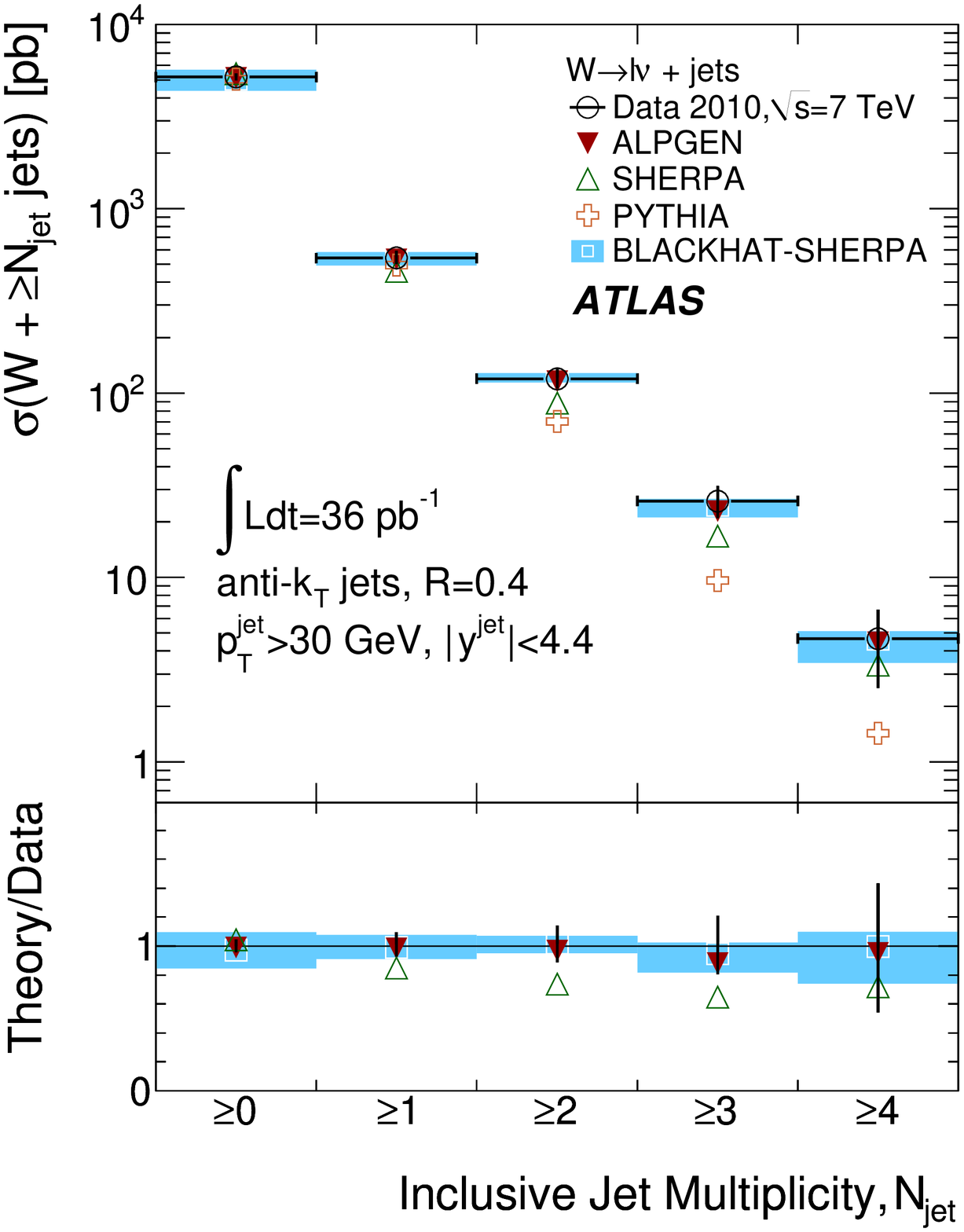,height=3.in}
\psfig{figure=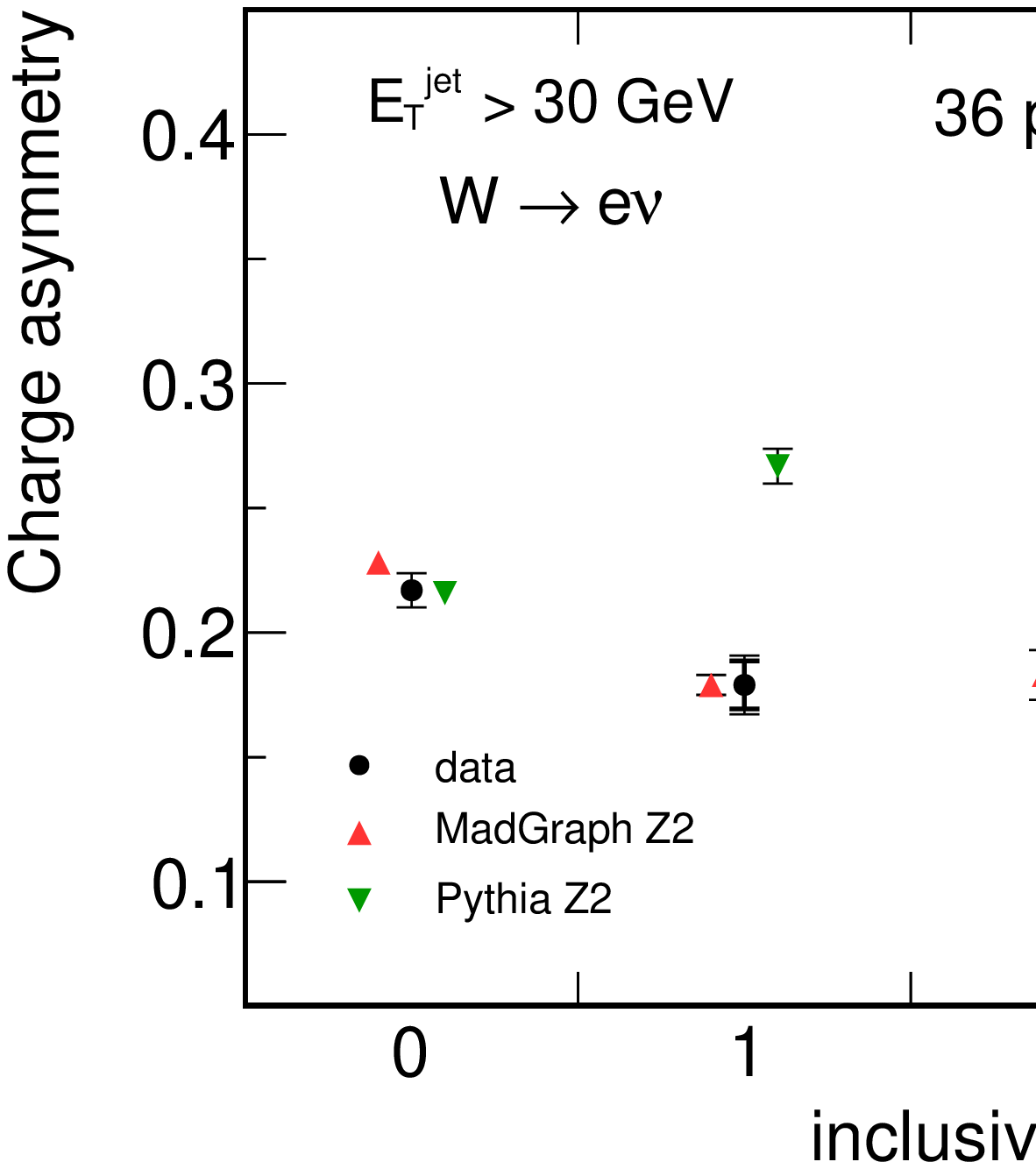,height=2.5in}
\caption{Left: The $W+n$-jets cross section, in inclusive jet multiplicity bins, measured by ATLAS~\protect\cite{atlas_w_jets}. Right: the lepton charge asymmetry in jet multiplicity bins for $W$ events, as measured by CMS~\protect\cite{cms_wz_jets}. \label{fig:wjets_rate}
}
\end{figure}

\begin{figure}
\psfig{figure=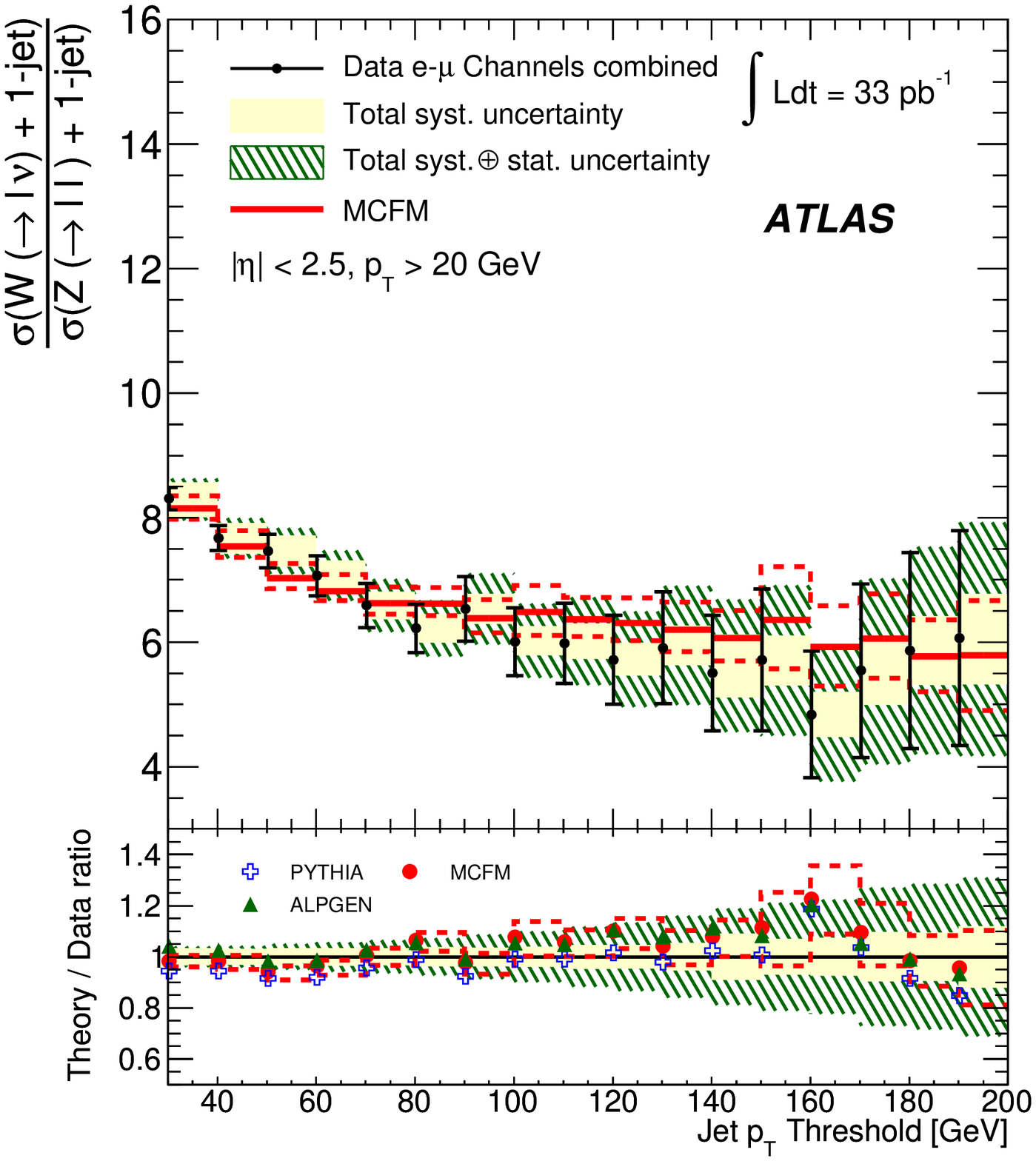,height=3.3in}
\psfig{figure=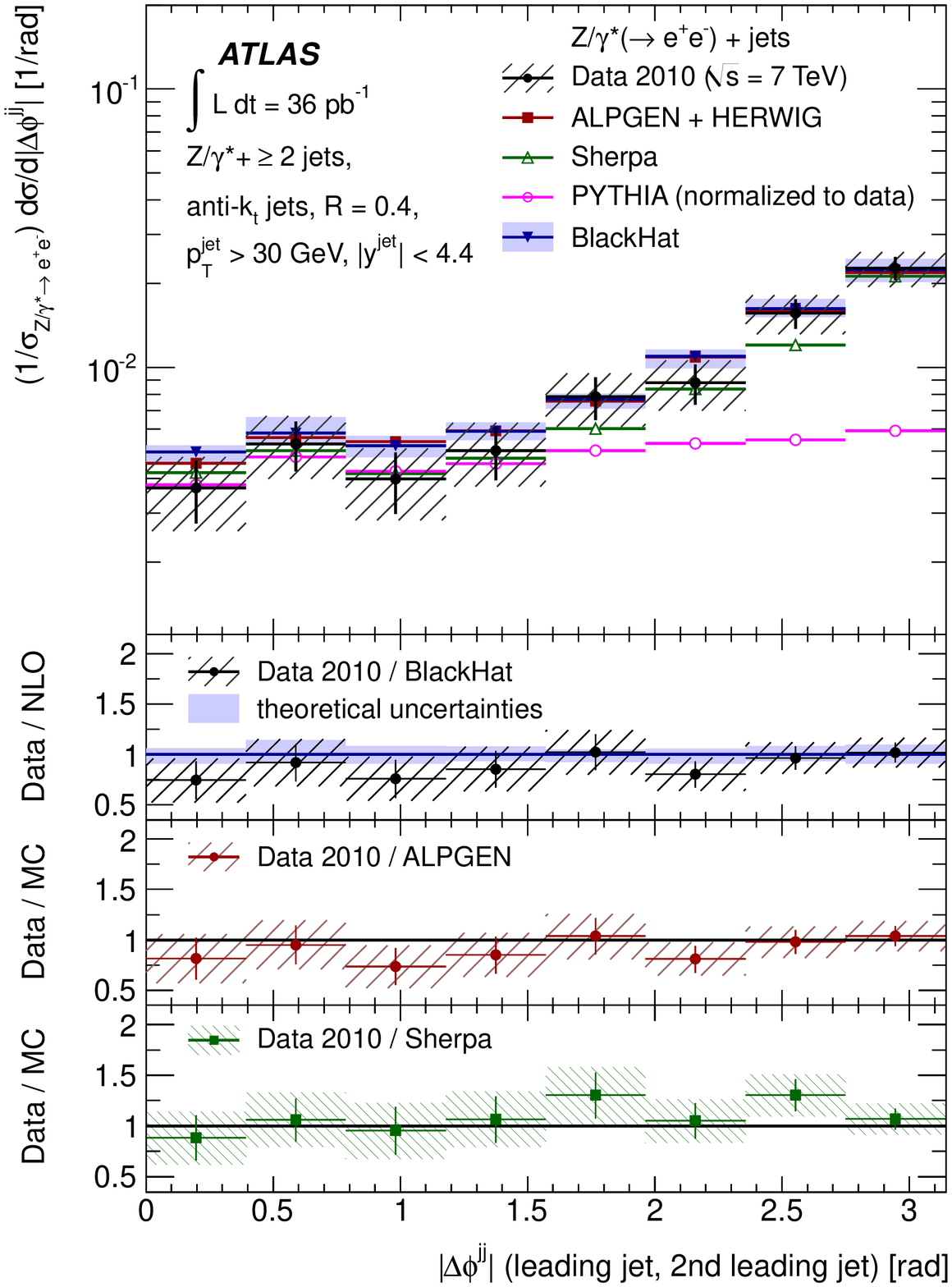,height=3.2in}
\caption{Left: the $W$+jet/$Z$+jet ratio as a function of the jet \pt threshold~\protect\cite{atlas_wz_ratio}. Right: $\Delta\phi$(jet,jet) in $Z$+$\ge2$~jet events~\protect\cite{atlas_z_jets}. Both measured by ATLAS.}
\label{fig:atlas_ratios}
\end{figure}

The first benchmark is to measure the inclusive jet rates produced in association with the $W$ or $Z$ 
(see Fig.~\ref{fig:wjets_rate})~\cite{atlas_w_jets}~\cite{atlas_z_jets}~\cite{cms_wz_jets}. Both 
experiments find the predictions of ALPGEN and SHERPA, and 
the latest NLO predictions from BLACKHAT, provide a good description of the data, within uncertainties.
The data uncertainties are dominated by the uncertainty on the jet energy scale. This, along with
some other uncertainties, can be partially cancelled by taking ratios, such as 
$W+n$-jets/$W+(n-1)$-jets, as measured at CMS~\cite{cms_wz_jets}.
ATLAS measure also the ratio of $W$+jet/$Z$+jet as a function of the  jet \pt\ threshold 
(see Fig.~\ref{fig:atlas_ratios})~\cite{atlas_wz_ratio}, which benefits from this cancellation while also testing 
the evolution of the predictions with increasing scale, and being sensitive to any new physics appearing
preferentially in one of the $W$ or $Z$ channels.
CMS also measure the $W$ charge asymmetry ($A_W=\frac{\sigma(W^+)-\sigma(W^-)}{\sigma(W^+)+\sigma(W^-)}$) in bins of inclusive jet multiplicity~\cite{cms_wz_jets} 
(see Fig.~\ref{fig:wjets_rate}). The data show a trend for reduced charge asymmetry at higher jet 
multiplicity, possibly due to the increased importance of gluon instead of valence quark initial 
states. This trend is reproduced in event generators which include explicit matrix elements for 
multiple jet production, but not in PYTHIA which relies on the parton shower to produce multiple jets.
ATLAS also measure a number of differential distributions in V+jet production, from 
individual jet momenta and rapidity ($y$) distributions, to correlations between jets and the boson, 
such as $\Delta y$(lepton, jet), $\Delta y$(jet, jet), dijet mass distributions in different jet
bins. These distributions pick out many different aspects of the underlying physics. For example,
the azimuthal angular separation, $\Delta\phi$(jet, jet), (see Fig.~\ref{fig:atlas_ratios}) highlighting the failure 
of the parton shower only approach in PYTHIA to produce well separated jets, and is also sensitive 
to multiple hard parton interactions producing a separate balanced (back-to-back) jet system in 
association with the $Z$.

\section{V + Heavy Flavour Jets}
\begin{figure}
\psfig{figure=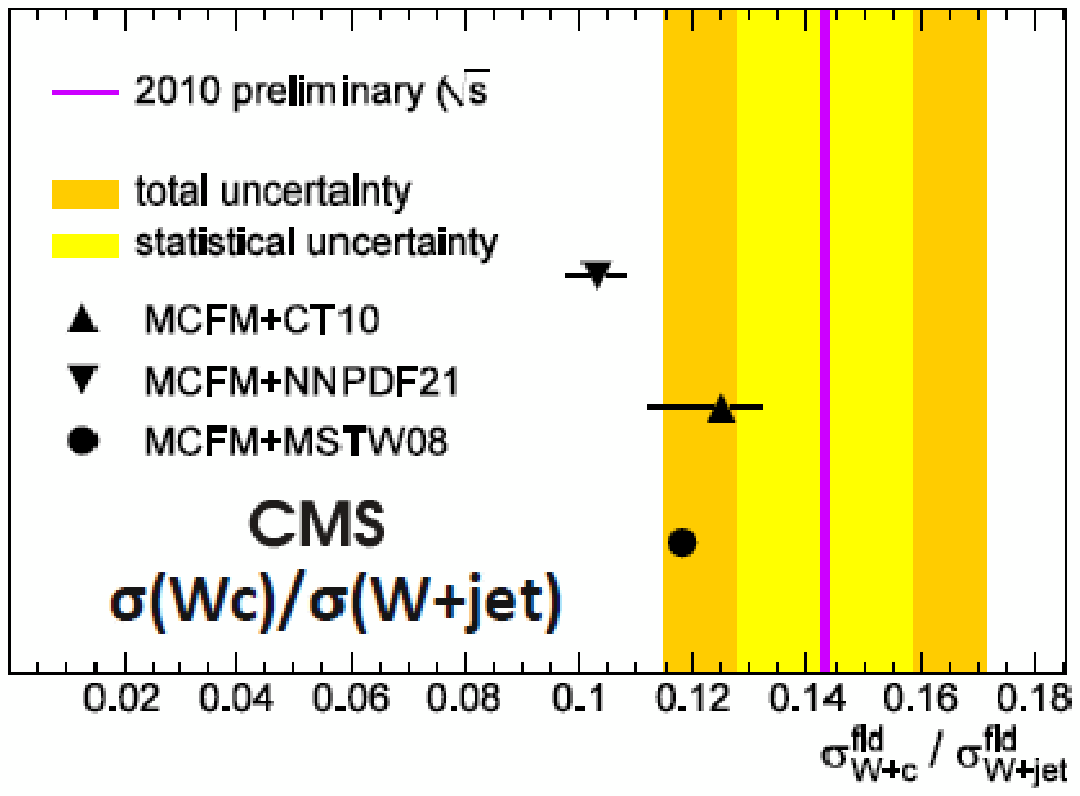,height=1.8in}
\psfig{figure=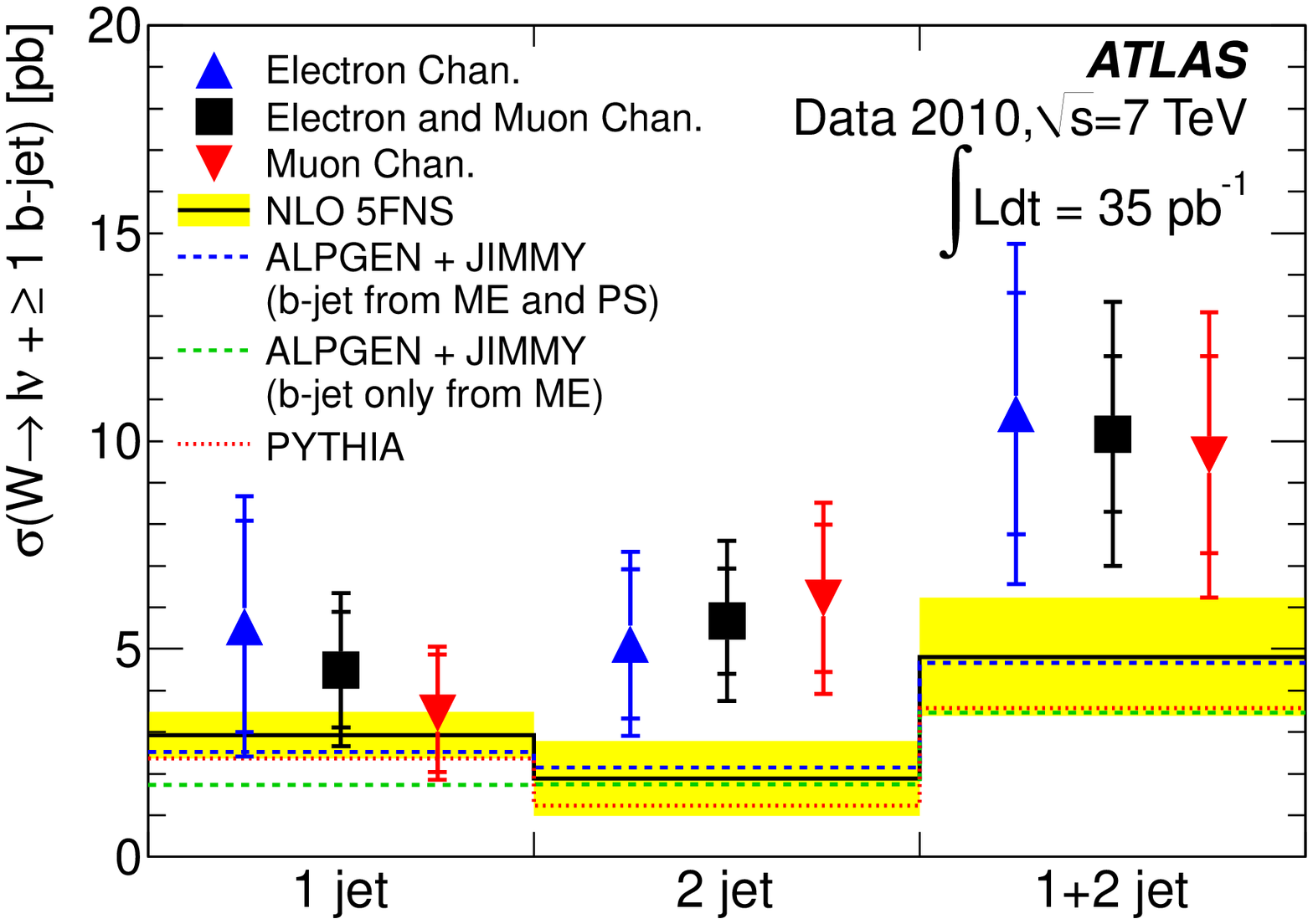,width=2.6in}
\caption{Left: constraints on the strange PDF from the CMS $W$+charm analysis~\protect\cite{cms_wcharm}. Right: ATLAS measurements of $W+b$-jet production~\protect\cite{atlas_wb}.\label{fig:Vhf}}
\end{figure}

Further information on the underlying physics can be obtained by identifying the flavour of
hadrons produced within jets. Measuring the production of $W$+charm, for example, gives a unique 
constraint on the strange quark content of the proton. Looking for semi-leptonic charm decays 
producing a muon inside a jet, and using the fact that in $W$+charm this muon will be of the 
opposite sign to the $W$ (compared to a random sign in most sources of background), CMS made a 
preliminary analysis of the 2010 dataset~\cite{cms_wcharm} to place constraints on the strange content of recent PDF sets~\cite{ct10}~\cite{nnpdf}~\cite{mstw} (see Fig.~\ref{fig:Vhf}). 

The production of $b$ hadrons  can be calculated in a 5-flavour PDF scheme (with a massless $b$-quark)
 or a 4-flavour scheme (including the $b$-mass); both suffer large theoretical uncertainties.
As V+$b$-jets forms the main background to low mass associated VH production, it is of particular
interest to constrain this signal. 
Jets containing $b$ hadrons can be identified based on the relatively long lifetime of these 
hadrons, and both experiments ``tag'' jets by looking for matching secondary decay verticies or 
tracks displaced from the primary vertex. This information, along with other 
sensitive quantities, are combined into multi-variate discriminants to maximise the tagging efficiency.
 The actual $b$-jet yield is separated from light flavour and charm jet mis-tags
 by fitting simulated templates to the data in discriminating variable such as the mass of the 
reconstructed secondary vertex in the event.
The production of $W+b$ suffers particularly large backgrounds from $W$+charm and top pair production.
However, ATLAS extract the $b$-jet yield in $W$ + 1- and 2-jet events, and see a 1.5$\sigma$ 
excess over NLO predictions~\cite{atlas_wb}  (see Fig.~\ref{fig:Vhf}).
The $Z+b$ case has much lower backgrounds, and ATLAS and CMS extract cross sections using different 
kinematic selections: ATLAS $3.55^{+0.82}_{-0.74} (\mathrm{stat})^{+0.73}_{-0.55} (\mathrm{syst})\pm 0.12 (\mathrm{lumi})$ pb (using 36~pb$^{-1}$)~\cite{atlas_zb} and CMS $5.84 \pm 0.08\,(\mathrm{stat.}) \pm 0.72 \,(\mathrm{syst.})
  _{-0.55}^{+0.25} \,(\mathrm{theory})$\,pb (using 2.2~fb$^{-1}$)~\cite{cms_zb}.
Both are compatible with NLO predictions within uncertainties.
Using the full 2011 dataset, CMS have also made a first preliminary study of the angular correlation
between $b$-hadrons in $Z$ events~\cite{cms_zbb}.
V+heavy flavour studies will clearly benefit significantly from the statistics available in  
the large datasets now collected by both ATLAS and CMS, and this will continue to be a strong area
for development.

\section{Conclusions}
Thanks to the energy reach of the LHC, it is already possible to make an
impressive range of measurements in the V+jets final state. So far, the latest theoretical 
predictions have generally been successful in describing the data, but as the precision increases 
with future results this remains an interesting area in which to test our understanding of such complex 
final states. As the understanding of these final states is essential also to describe these
backgrounds to searches for new physics like the Higgs boson, it also remains an essential area for
development on both the experimental and theoretical side.


\section*{Acknowledgements}
The author wishes to grateful to the Royal Society for supporting this work.

\end{document}